\documentclass[12pt,preprint]{aastex}

\newcommand{\Mdot}[2]{\mbox{${#1}\times10^{-{#2}}$\,M$_\odot$~yr$^{-1}$}}
\newcommand{\Msun}{\mbox{\,M$_\odot$}}
\newcommand{\Lsun}{\mbox{\,L$_\odot$}}
\newcommand{\vunit}{\mbox{\,km\,s$^{-1}$}}
\newcommand{\mic}{\mbox{$\,\mu$m}}
\newcommand{\pion}[2]{{#1}\,{\sc {#2}}}

\newcommand{\gtsimeq}{\raisebox{-0.6ex}{$\,\stackrel
	{\raisebox{-.2ex}{$\textstyle >$}}{\sim}\,$}}

\newcommand{\rs}{RS~Oph}
\newcommand{\sirtf}{\mbox{\it Spitzer}}

\shorttitle{Silicate dust in RS~Oph}
\shortauthors{Evans et al.}

\begin{document}


\title{Silicate dust in the environment of RS~Ophiuchi following
the 2006 eruption}

\author{A. Evans\altaffilmark{1},
C. E. Woodward\altaffilmark{2},
L. A. Helton\altaffilmark{2},
J. Th. van Loon\altaffilmark{1},
R. K. Barry\altaffilmark{3},
M. F. Bode\altaffilmark{4},
R. J. Davis\altaffilmark{5},
J. J. Drake\altaffilmark{6},
S. P. S. Eyres\altaffilmark{7},
T. R. Geballe\altaffilmark{8},
R. D. Gehrz\altaffilmark{2},
T. Kerr\altaffilmark{9},
J. Krautter\altaffilmark{10},
D. K. Lynch\altaffilmark{11},
J.-U. Ness\altaffilmark{12},
T. J. O'Brien\altaffilmark{5},
J. P. Osborne\altaffilmark{13},
K. L. Page\altaffilmark{13},
R. J. Rudy\altaffilmark{11},
R. W. Russell\altaffilmark{11},
G. Schwarz\altaffilmark{14},
S. Starrfield\altaffilmark{12},
V. H. Tyne\altaffilmark{1}
}

\altaffiltext{1}{Astrophysics Group, Lennard-Jones Laboratories, Keele
University, Staffordshire, ST5 5BG, UK,
{\it ae@astro.keele.ac.uk, jacco@astro.keele.ac.uk, vht@astro.keele.ac.uk}}

\altaffiltext{2}{Department of Astronomy, School of Physics \& Astronomy, 116 Church
      Street S.E., University of Minnesota, Minneapolis, MN 55455, USA,
      {\it ahelton@astro.umn.edu, chelsea@astro.umn.edu, gehrz@astro.umn.edu}}

\altaffiltext{3}{Exoplanets and Stellar Astrophysics Laboratory, NASA Goddard
Space Flight Center Greenbelt, MD 20771, USA
{\it richard.k.barry@nasa.gov}}

\altaffiltext{4}{Astrophysics Research Institute, Liverpool John Moores University,
      Twelve Quays House, Birkenhead CH41 1LD, UK,
{\it mfb@astro.livjm.ac.uk}}

\altaffiltext{5}{Jodrell Bank Centre for Astrophysics, University of Manchester,
      M13 9PL, UK {\it tob@jb.man.ac.uk, rjd@jb.man.ac.uk}}

\altaffiltext{6}{Harvard-Smithsonian Center for Astrophysics (CfA), 60 Garden
      Street, Cambridge, MA 02138, USA. {\it jdrake@cfa.harvard.edu}}

\altaffiltext{7}{Centre for Astrophysics, University of Central Lancashire,
      Preston, PR1 2HE, UK, {\it spseyres@uclan.ac.uk}}

\altaffiltext{8}{Gemini Observatory, 670 N. A'ohoku Place, Hilo, HI\,96720, USA,
{\it tgeballe@gemini.edu}}

\altaffiltext{9}{Joint Astronomy Centre, 660 N. A'ohoku Place, University Park,
      Hilo, Hawaii 96720, USA, {\it t.kerr@jach.hawaii.edu}}

\altaffiltext{10}{Landessternwarte, K\"{o}nigstuhl, D-69117 Heidelberg, Germany,
     {\it j.krautter@lsw.uni-heidelberg.de}}

\altaffiltext{11}{The Aerospace Corporation, Mail Stop M2-266, P.O. Box 92957, Los
      Angeles, CA~90009-2957, USA,
{\it David.K.Lynch@aero.org, Richard.J.Rudy@aero.org, Ray.Russell@aero.org}}

\altaffiltext{12}{School of Earth \& Space Exploration, Arizona State University,
      PO Box 871404, AZ 85287-1404, USA,
 {\it sumner.starrfield@asu.edu, ness@susie.la.asu.edu}}

\altaffiltext{13}{Department of Physics and Astronomy, University of Leicester,
      Leicester, LE1 7RH, UK, {\it julo@star.le.ac.uk, kpa@star.le.ac.uk}}

\altaffiltext{14}{Steward Observatory, University of Arizona, 933 North Cherry
      Avenue, Tucson, AZ 85721, USA, {\it gschwarz@as.arizona.edu}}



\begin{abstract}
We present further \sirtf\ Space Telescope observations of the recurrent nova
{\rs}iuchi, obtained over the period 208--430~days after the 2006 eruption.
The later \sirtf\ IRS data show that the line emission and free-free
continuum emission reported earlier is declining, revealing 
incontrovertible evidence for the presence of
silicate emission features at 9.7 and 18\mic. We conclude that the silicate
dust survives the hard radiation impulse and shock blast wave from the eruption.
The existence of the extant dust may have significant implications for
understanding the propagation of shocks through the red giant wind and likely wind
geometry.
\end{abstract}



\section{Introduction}

\rs\ is a recurrent nova (RN) that erupted in 1898, 1933, 1958, 1967, 1985
\citep{wallerstein}. It consists of a
semi-detached binary (orbital period 455.7~days) comprising a
roche-lobe-filling red giant (M2III) and a massive ($\gtsimeq1.2\Msun$) white dwarf
\citep[WD;][]{fekel}. The eruption follows a thermonuclear runaway on the
surface of the WD \citep{tnr}. In the case of the \rs\ class of RN, however, the
ejected material runs into, and shocks, a dense red giant (RG) wind
\citep{bk85,obrien92}.

The 1985 eruption was, for the first time, the subject of a
multi-wavelength observational campaign, from the radio to the X-ray
\citep{vnu}. Its most recent eruption, on 2006 February 12.83
\citep[][ we take this to define the origin of time post-outburst]{hirosawa},
was the subject of an even more intensive
observational campaign. Infrared (IR) observations
\citep{das06,evans07a,evans07b}  showed evidence for the shock, seen also at
radio \citep{eyres,obrien} and X-ray \citep{bode06,sokoloski,ness,osborne}
wavelengths, as the ejecta interacted with the RG wind.

Observations of \rs\ obtained with the Spitzer Space Telescope
\citep[{\it Spitzer};][]{werner, gehrz07} during the period from 2006 April 16 
through 26~UT ($\simeq 67$ days after outburst) were described by 
\cite{evans07b}. We present here further IR observations conducted as 
part of a long duration synoptic campaign, obtained later in the outburst 
with the \sirtf\ Infrared Spectrometer \citep[IRS;][]{houck}.

\section{Observations} 

\rs\ was observed with the \sirtf\ IRS as part of the Director's Discretionary
Time program, Program Identification (PID)~270 on 
2006 September 9.83~UT (day 208), 2006 October 10.4~UT (day 246) and 2007 
April 19.5~UT (day 367.2). Observations were
performed using all IRS modules and the blue peak-up on \rs. The spectroscopy
consisted of 5 cycles of 14 second ramps in short-low mode, 5 cycles of 30
second ramps in both short-high and long-low modes, and 5 cycles of 60 second
ramps in long-high mode. IRS basic calibrated data products (BCDs) were
processed with versions 14.4.0 (2006 observations) and 16.1.0 (2007 observation)
of the pipeline.
The extracted spectra were not defringed. Full details of the data reduction
process are given in \cite{evans07b} and are not repeated here. 

\begin{figure}
\includegraphics[angle=-90,scale=.40]{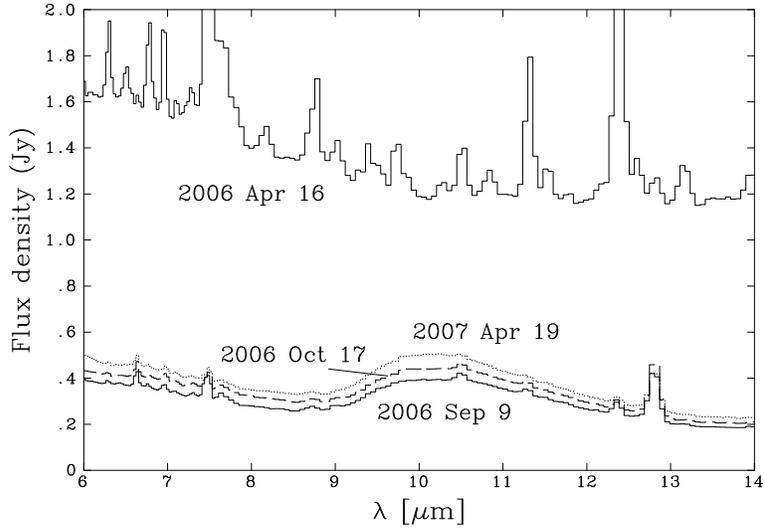}
\caption{\sirtf\ IRS spectra of \rs\ over the period 2006 April (hot gas
dominant) to late 2006 and later (silicate features prominent).}
\label{evol}
\end{figure}

Fig.~\ref{evol} shows, with the same flux density scale and over the wavelength
range 6--14\mic, the \sirtf\ spectrum from \cite{evans07b}, when free-free
radiation and fine structure lines dominated the emission, along with the
three new spectra reported in this paper. In the period covered by the
observations reported here there seems to
have been little change in the level of the emission: the differences of
$\sim10$\% are comparable with the uncertainty in the flux calibration.
We conclude that, in 2006 April,
emission from the hot gas dominated the dust emission but as the 
emission from the gas subsided, emission from dust was revealed. The low 
resolution \sirtf\ spectrum of 2006 Sept. 09.83~UT is shown in Fig.~2, in 
which the silicate features at 9.7\mic\ and 18\mic\ are clearly 
visible.

We confine our discussion in this paper to the silicate features, and defer a
discussion of the evolving line and (gas) continuum emission to a later paper.

\section{Discussion}

The formation of silicate dust requires an environment in which
$\mbox{O}>\mbox{C}$ by number \citep{whittet}. Therefore on the assumption that
the silicate dust we see in the \sirtf\ spectrum arises in the RG wind
(see below), we conclude that the RG in the \rs\ system cannot be
carbon-rich; this is consistent with the presence of
TiO bands in the optical \citep{blair}, and the presence of deep \pion{O}{i}
absorption combined with no carbon absorption in the X-ray range \citep{ness}.

\rs\ is listed as a 12\mic\ source in the IRAS Point Source Catalog, with a flux
density (not colour corrected) 0.42~Jy and upper limits in the other IRAS bands.
To obtain a better estimate of the mid-IR flux densities, we retrieved the
original scans from the IRAS database \citep{iras} and used the
GIPSY software to measure flux densities from traces through the position
of the star. In this way the 12\mic\ value is refined to $0.37\pm0.02$~Jy,
and a marginal detection, $0.09\pm0.03$~Jy, is recovered at 25\mic; \rs\ was
not detected at 60 or 100\mic, to uninterestingly large upper limits. These
data, along with the 10\mic\ flux from \cite{geisel}, the 2MASS
survey\footnote{The 2MASS survey is a joint project of the University of
Massachusetts and the Infrared Processing and Analysis Center/California
Institute of Technology, funded by NASA and the NSF.}
\citep[][ data obtained on 22 April 1999]{2mass} and the \sirtf\
spectrum from 2006 September, are shown in Fig.~\ref{bunny}.

Given that the IRAS, 2MASS and \cite{geisel} data were obtained between
eruptions we are confident that they represent the quiescent state of \rs.
Moreover the fact that the \sirtf\ data are superficially consistent with
the inter-outburst IR observations (see Fig.~\ref{bunny}), and that the
dust was present at day 430, suggests that the dust seen by \sirtf\ is 
likely present between outbursts (note that \cite{hodge} found no evidence
of dust in ISO spectra of \rs, but their data only went as far as 9\mic).
We have made a preliminary fit using {\sc dusty} \citep{dusty} with a 3600~K
blackbody (to represent the RG during quiescence) as input radiation field.
Using silicate optical constants from \cite{ossenkopf}, we find optical depth
$\tau_{\rm V} \simeq 0.1$ ($\pm\sim10$\%), dust temperature at the inner edge of
the dust shell $T_0 \simeq 600$~K ($\pm\sim100$~K) and dust shell inner radius
of $1.5 \times 10^{14}$~cm. We derive a luminosity $L_* \simeq 440$\Lsun,
mass-loss rate $\dot{M} \simeq \Mdot{2.3}{8}$ \citep[typical for these stellar
parameters;][]{vanloon} for a dust to gas mass ratio of 0.005, and a terminal
wind speed $v_\infty=8.1$\vunit.

While we caution against taking these derived parameters too literally, the 
`outer' edge of the wind (defined by $v_\infty\Delta t$, where $\Delta t$
is the time between the 1985 and 2006 eruptions) is $\simeq5.4 \times
10^{14}$~cm. The ejecta from the 2006 eruption, travelling at constant
velocity, would have overtaken the outer edge of the wind even at the earliest
of our observations ($t = 208$~days). However the shock, which is
decelerating \citep{bode06, das06, evans07a} may not, as yet, have reached the
wind edge. Taking $\tau_{\rm V}$/$\tau_{9.7} \simeq 17.8$
\citep[cf.][]{mathis} and silicate opacity from \cite{ossenkopf}, we deduce a
dust mass of $4.1 \times 10^{-9}$~M$_{\odot}$. Also, the inner radius
of the dust shell is rather greater than the binary separation
($\simeq2.2\times10^{13}$~cm, assuming that the mass of the \rs\ system is 
$\simeq2$\Msun).

We have determined the full-width at half-maximum (FWHM) of the 9.7\mic\
feature by drawing a linear continuum between 8.5\mic\ and 12.5\mic\
(see Fig.~\ref{9.7}, in which the fluxes at 10\mic\ have been normalized to the
2006 September value). We find that, on  2006 September 9 and October 17, the
FWHM was 2.17\mic\ and 2.14\mic\ respectively, with no evidence for change;
indeed it is difficult to distinguish the silicate features for these two epochs.
By 2007 April 19, however, the FWHM was significantly
smaller, 1.99\mic, perhaps because the dust had experienced some
processing (resulting in a change in the dust optical properties) since
2006 September/October.
The profile of the 9.7\mic\ feature in \rs\ broadly resembles those of the
``Type~C'' silicate features of AGB stars in  \citeauthor{speck}'s
(\citeyear{speck}) classification. 
However the silicate feature in \rs\ may dffer in detail from those of AGB stars
for several reasons, such as lack of hot dust in \rs, or to grain size,
shape, and composition differences. Differences may also arise because the dust
in \rs\ is persistently exposed to ultra-violet (UV) radiation from the WD, as well 
as suffering frequent exposure to hot ionized gas (see below); the latter
is known to anneal amorphous silicate \citep{carrez}.


Dust formation in the winds of classical novae during eruption is
well-documented \citep{evans08,gehrz08}. Indeed our observations
were obtained during an inflection in the visible light curve, where
the $V$-band magnitude abruptly decreased by $\sim 0.6$~mag, recovering
shortly thereafter. In classical
novae such inflections, accompanied by changes in the IR spectral
energy distribution, are associated with dust condensation in
the ejecta.

However, in \rs\ we are confident that dust did not form in 
the ejecta, for two reasons. First, the temperature of the shocked gas
\citep[$\sim10^6$~K;][]{evans07a,evans07b} is not conducive to grain formation, 
which requires dense, relatively cool ($\simeq 2000$~K) gas. Second,
the strength of the 18\mic\ feature relative to that of the 9.7\mic\ 
feature indicates that the dust has been processed.
In freshly-condensed silicate dust, the 18\mic\ feature is expected
to be weak by comparison with the 9.7\mic\  
feature \citep{nuth}. Alternatively, the 18\mic/9.7\mic\ ratio for 
silicate features in these circumstellar environments could be solely dependent 
on dust temperature. We have already noted the apparent paucity of hot dust 
in \rs, so the 9.7\mic\  feature will be less strong. In fact our 
{\sc dusty} fit (Fig.~\ref{bunny}) over-estimates the 9.7~\mic\ feature
strength, even though the model satisfactorily fits the dust continuum as well
as the 18\mic\ feature. We conclude that the silicate emission evident in the
\rs\ \sirtf\ spectra is inconsistent with dust formation in the 2006 eruption.


Previous models of the evolution of the remnant of \rs\ following eruption
\citep[e.g.][]{bk85,obrien92} have, for reasons of simplicity, assumed
spherical symmetry. Accordingly the ejecta sweep
into the RG wind, are shocked, and eventually reach the outer edge of the wind.
In this scenario the entire RG wind is completely engulfed by the ejecta, and
is replenished only after the eruption has subsided and the
RG wind starts anew. As a result any dust condensing in the RG
wind would also be engulfed by the shocked gas.

However the presence of warm silicate dust so soon after the 2006 eruption, and
its likely presence between eruptions, may imply that a substantial portion
of the RG wind never `sees' the eruption at all. Assuming that the
RG fills its Roche lobe, its radius is $\simeq0.7$ of the binary separation for a
mass ratio (secondary/primary) $\simeq0.4$. It therefore presents a solid angle
$\sim0.3$~steradians to the WD (i.e. it covers $\sim2.5$\% of the WD's sky).

When the RN eruption occurs on the surface of the WD it sweeps past the RG in
$\sim 1.5$ days, potentially arriving at the nominal outer edge of the wind in a
few tens of  days. This is $\ll$ than the orbital period, so
$\sim2.5$\% of the wind is shielded by the RG and sees
neither the eruption itself nor the expanding ejecta. However the
situation is more complex than this in view of the radial-dependence of the
wind density, and it is likely that the shocked gas will
encroach into the shielded volume. Nevertheless our estimate of 2.5\% for the
fraction of shielded wind is likely of the right order: shielding of the wind
by the RG is unlikely to account for the dust we see.

On the other hand, the density of circumstellar material in binaries is
greatest in the binary plane \citep[e.g.][ and \cite{gehrz-ry} for the case
of the massive binary RY~Sct]{spruit}, so the
RG wind will not be spherically symmetric. Outflow, both of the wind and the
2006 ejecta, will be inhibited in the binary plane and indeed, anisotropies are
suggested by HST \citep{bode07} and radio 
\citep{obrien} observations, and IR interferometry \citep{monnier}.
We expect a significant amount of dust to survive in the binary plane;
such a dense medium would be consisent with the shock deceleration
seen in the early Swift \citep{bode06} and IR
\citep{das06,evans07a} data. 

The higher density material in the binary plane is effectively shielded
and may not experience the eruption (which in any case may not have been
spherically symmetric); this offers, in part, an explanation for the continued
presence of silicate dust in the environment of \rs\ after eruption. 


Much of the dust in the RG wind away from the binary plane would not survive the
surge of UV radiation emitted at outburst. The temperature of grains at the
inner edge of the dust shell would rise by a factor $[L_{\rm
Edd}/L_*]^{1/5}\simeq2.5$ when heated by UV radiation from a source at the
Eddington limit, $L_{\rm Edd}$, for a Chandrasekhar-mass WD. Thus grains at 600~K
at quiescence would be heated to $\simeq 1500$~K, in excess of the nominal
evaporation temperature \citep[$\simeq1300$~K;][]{speck} of silicate dust.
Indeed, silicate grains would not survive the eruption out to distance
$r_{\rm evap} \sim 2.2 \times 10^{14}$~cm from the 
WD in outburst. Furthermore, it is likely that a substantial amount of dust 
may have survived in denser regions swept-up in winds from previous
outbursts.

Any dust beyond the evaporation distance would survive the UV blast, so
we consider whether this dust would survive the passage of the shock.
As discussed in earlier papers \citep{evans07a,evans07b}, the gas
temperature implied by the IR fine structure and coronal lines likely ranges
from $\sim150\,000$~K to $\sim900\,000$~K, with an electron density
$\sim2.2\times10^5$~cm$^{-3}$ in the cooler region; the x-ray data
indicate shock temperatures even higher than this \citep{bode06}.
Dust grains in this environment will be sputtered as they are engulfed by the
shock. Assuming that the sputtering threshold for silicate grains is
$\sim6$~eV \citep{tielens}, the time-scale for complete erosion of 0.1\mic\
silicate grains in the environment implied by the IR data is
$\simeq{Y}^{-1}$~years, where the yield $Y$ (the number of grain atoms ejected
per incident ion) is $\sim10^{-3}$ \citep{tielens}. As this erosive time-scale is
$\gg{\Delta}t$ (the inter-outburst interval) any grains in the RG wind would
survive passage of the shock, but might well be processed during the passage of
the hot gas.

Our conclusion is consistent with $N$-band interferometric observations
conducted on day~3.8, which may have independently seen the
dust we report here \citep{bar07}. The nearest constructive peak in the
interferometer's acceptance pattern is at an angular distance from the central
null fringe that depends on wavelength as $\lambda/B$, where $B$ is the length
of the projected baseline (84~m). At the center of the $N$-band, in which
\pion{Si}{i}~9.407\mic\ and dust are detected, the peak transmission is at
angular separation $\simeq12.5$~mas from the center of the source brightness
distribution, equivalent to $\simeq3\times10^{14}$~cm at a distance of 1.6~kpc.
The initial shock velocity would need to have been $\sim9000$\vunit\ to reach
this distance, travelling at constant velocity, far higher than observed
\citep[e.g.][]{evans07a,bode07}. Therefore this emission line was
not excited by passage of the shock front but by the UV flash from the TNR
event. This measurement is consistent with our assertion that dust detected by
\sirtf\ was not created by the collision of the ejecta with the RG wind 
or by subsequent condensation, but was more likely formed during quiescence.

Our continuing \sirtf\ observations of \rs\ will enable us to examine the
effects of irradiation on the dust, and of the re-formation of dust as the
circumbinary environment is re-established by RG wind infill. 

\section{Concluding remarks}

We have presented further \sirtf\ IRS observations of \rs\ in the aftermath
of its 2006 eruption. Unlike our earlier observation, which showed emission
by the hot shocked gas, more recent IR spectra reveal the presence of
silicate dust.

The dust we see could not have been formed in the 2006 eruption. It 
most probably survived the 2006 eruption, because a portion of
the RG wind (including the dust) did not see the eruption at all, and
because the passing shock failed to destroy any dust that survived the UV
blast at the eruption. This conclusion may have major implications for the
evolution of the shock, and of the long-term survival of the RG wind
between eruptions. We predict that the dust we  see
around \rs\ will persist until the next eruption.

\acknowledgments

This work is based on observations made with the \sirtf\ Space
Telescope, which is operated by the Jet Propulsion Laboratory, California
Institute of Technology under a contract with NASA; we acknowledge the
award of Director's Discretionary Time.
LAH, CEW and RDG are supported in part by NASA through contracts 1256406,
1215746 and 1267992 issued by JPL/Caltech to the University of Minnesota.
TRG is supported by the Gemini Observatory, which is operated by the Association
of Universities for Research in Astronomy, Inc., on behalf of the international
Gemini partnership of Argentina, Australia, Brazil, Canada, Chile, the United
Kingdom, and the United States of America.
The work of DKL, RJR and RWR is supported by The Aerospace
Corporation's Independent Research and Development Program.
MFB was supported by a PPARC Senior Fellowship.
J.-U. N. was supported NASA through a
Chandra Postdoctoral Fellowship grant PF5-60039 awarded by the Chandra
X-ray Center, which is operated by the Smithsonian Astrophysical
Observatory for NASA under contract NAS8-03060.
JPO and KLP acknowledge support from PPARC.
SS acknowledges partial support from NSF and NASA grants to Arizona State
University.

{\it Facilities:} \facility{\sirtf\ (IRS)}

\clearpage

\begin{figure}
\includegraphics[angle=-90,scale=.40]{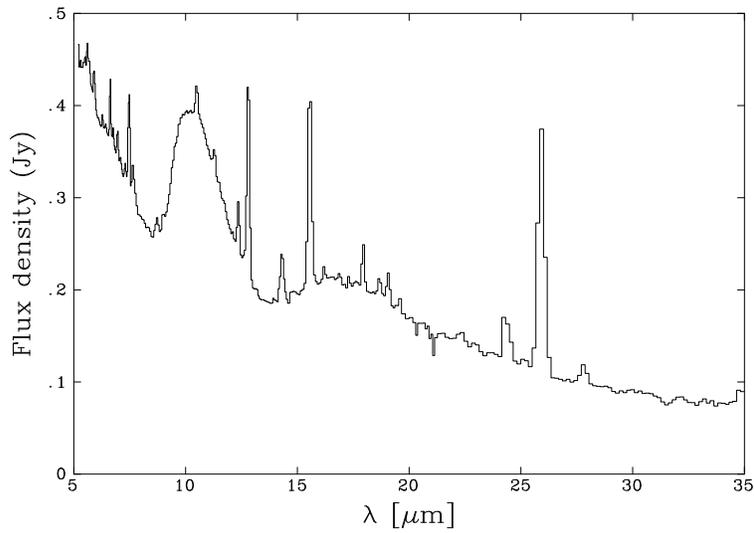}
\caption{\sirtf\ low resolution IRS spectrum of \rs\ obtained on 2006 September
9.8~UT. The silicate features at 9.7 and 18\mic\ are clearly visible.} 
\label{rsoph_dust}
\end{figure}

\begin{figure}
\includegraphics[angle=0,scale=.40]{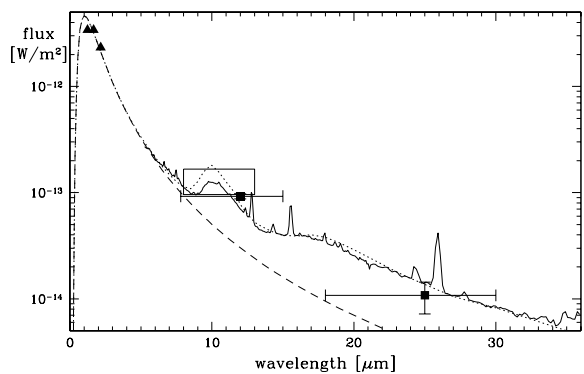}
\caption{\sirtf\ IRS spectrum from 2006 September 9.8~UT, together 
with data from 2MASS, IRAS, and \citet{geisel}. Triangles: 2MASS; 
squares with error bars: IRAS; large open square: \cite{geisel}.
The broken curve is a 3600~K blackbody, used as input into {\sc dusty}; the
dotted curve is the {\sc dusty} fit. For details see \S 3.}
\label{bunny}
\end{figure}

\begin{figure}
\includegraphics[angle=-90,scale=.40]{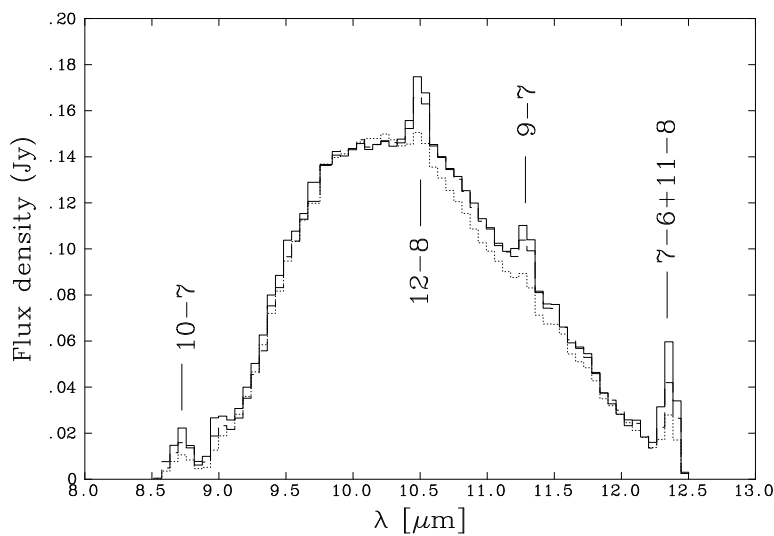}
\caption{\sirtf\ spectrum of \rs\ for 2006--2007; a linear continuum has been
removed from the 8.5--12.5\mic\ region. Principal H
recombination lines ($n-m$) are identified. Spectra for 2006 September and
October are essentially identical; note that the 9.7\mic\ feature for 2007 April
(dotted line) is noticeably narrower.}
\label{9.7}
\end{figure}

\end{document}